\documentstyle[graphicx, 12pt]{article}
\begin{document}

\small
\hoffset=-1truecm
\voffset=-2truecm
\title{\bf The circular loop equation of a cosmic string with time-varying tension in de Sitter spacetimes}
\author{Yunqi Liu \hspace {1cm} Hongbo Cheng\footnote {E-mail address: hbcheng@public4.sta.net.cn}\\
Department of Physics, East China University of Science and
Technology,\\ Shanghai 200237, China}

\date{}
\maketitle

\begin{abstract}
In this work the equation of circular loops of cosmic string
possessing time-dependent tension is studied in the de Sitter
spacetime. We find that the cosmic string loops with initial
radius $r(t_{0})>0.707L$, L de Sitter radius, should not collapse
to form a black holes. It is also found that in the case of
$r(t_{0})<0.707L$ a loop of cosmic string whose tension depends on
some power of cosmic time can not become a black hole if the power
is lower than a critical value which is associated with the
initial size of the loop.
\end{abstract}

\vspace{8cm}
\hspace{1cm} PACS number(s): 98.80.Cq

\newpage
As a kind of topological defects, cosmic strings including their
formation, evolution and observational effects have been a focus
for more than twenty years [1-4]. The model was thought as seed of
large scale structure in our universe. Although the improved
observational data like CMB anisotropy observations and the WMAP
experiments do not favour cosmic strings as the origin of the
primordial density perturbations giving rise to the existence of
galaxies and clusters due to limits on their tension
($G\mu\leq10^{-6}$) [5-7], cosmic strings have several potentially
important and distinct astrophysical features. The topological
defects, including cosmic strings, were inevitably formed at the
end of both conventional inflation [1-3] and brane inflation [8,
9], so we have to face and get to know this kind of model.
Furthermore cosmic strings still have strong influence on various
astrophysics such as gravitational lensing effects [10, 11],
gravitational wave background [12, 13], early reionization [14,
15] and so on. The model of cosmic string can be used to explain
some important astronomical phenomena powerfully. For example, a
gravitational lens called CSL-1 invoking two imagines of
comparable magnitude of the same giant elliptical galaxy were
discovered. It is interesting that many similar objects were found
in the vicinity [16, 17].

It is necessary to investigate the evolution of cosmic string
loops carefully. Once the cosmic strings formed, they were not
static under their own force of tension but envolved instead.
Therefore cosmic strings will produce loops unavoidably. The
strings can collide and intersect to undergo reconnections. The
reconnections of long strings and large loops will generate small
loops copiously. In general, string networks consist of long
strings and closed string loops. There are observational results
which support the existence of cosmic string loops in our Universe
[11, 18]. Schild et al. discovered and analyzed the anomalous
brightness fluctuations in the multiple-image lens system
Q0957+561A, B, which has been studied intensively for 25 years
[11, 18]. The system with brightness fluctuations consists of two
quasar images separated by approximately $6''$. They thought that
the observational results were known to be images of the same
quasar not only because of the spectroscopic match, but also
because the images fluctuate in brightness, and the time delay
between fluctuations is always the same. They suggested that the
effect may be due to lensing by an oscillating loop of cosmic
string between us and the lensing system, because cosmic string
loops supply a quantitative explanation of such synchronous
variations in two images. However the fate of cosmic string loops
is not entirely satisfactory. According to the past discussions
nearly all loops will become black holes except in the special
cases. In Minkowski spacetime and Robertson-Walker universe, the
loops will collapse to form black holes under their own tension
certainly instead of remaining oscillating loops after the loops
formation [19, 20]. In de Sitter backgrounds, only loops with
large initial radii can avoid becoming black holes [19, 21, 22].
However a large loop will evolve to be a lot of smaller loops,
loops can not live in a de Sitter spacetime unless they are very
large. Clearly in the environment with a positive cosmological
constant few cosmic string loops can survive. Here we must point
out that black holes are final result for loops of cosmic string
in our universe bases on the hypothesis that the tension of cosmic
string is constant. It may be also worth pointing out that the
evolution of loops of different kinds of cosmic strings needs to
be investigated.

It is really fundamental to explore the evolution of loops of
cosmic string with time-dependent tension. In fact the cosmic
strings with time-varying tension appeared generally in
cosmological situations. So far in nearly all researches, that the
tensions of cosmic string are constant is just an assumption. M.
Yamaguchi put forward the important issue that the tensions of
cosmic strings can depend on the cosmic time [23]. Some researches
on the problem were performed and the significant and interesting
conclusions were drawn [23, 24]. The tensions of cosmic strings
can depend on the cosmic time. For example, a potential containing
one complex scalar field $\phi$ and one real scalar field $\chi$
can be written as $V(\phi,
\chi)=\frac{\lambda}{4}(|\phi|^{2}-\chi^{2})^{2}+\frac{1}{2}m_{\chi}^{2}\chi^{2}$.
The backreaction to the oscillation of field $\chi$ is negligible
as the coupling constant $\lambda$ is sufficiently small. The
tension of string $\mu$ is associated with the root mean square of
the expection value of field $\chi$ and can be denoted as
$\mu\propto a^{-3}$ where $a$ is the scale factor which is
proportional to $t^{\frac{1}{2}}$ in the radiation-dominated era
and $t^{\frac{2}{3}}$ in the matter-dominated era [23]. In the
case where the tension depends on the power of the cosmic time
like $\mu\propto t^{q}$, such cosmic strings go into the scaling
solution when $q<1$ in the radiation domination and
$q<\frac{2}{3}$ in the matter domination respectively. The authors
of [23, 24] also show that the CMB and matter power spectra
induced by cosmic strings with changeable tension can be different
significantly from those generated by the conventional cosmic
strings with constant tension, which means that a lot of related
topics are mysterious and need to be explored. Until now little
contribution is made to scrutinize the evolution of cosmic string
loops possessing time-dependent tension.

Here we are going to obtain the equation of circular loops of
cosmic string with time-varying tension in the de Sitter spacetime
due to our accelerating universe. We wonder the time dependence of
the tension on the evolution and fate of the cosmic string loops.
In the case of constant tension loops in de Sitter backgrounds
will keep on expanding if they obey the necessary conditions that
their initial radii are larger than $0.707L$, $L$ de Sitter radius
[21]. Now as the first step we focus on the models with cosmic
time power tension. When the tension of cosmic string is
proportional to power of cosmic time, we search the new necessary
conditions leading the loops to keep enlarging. In this paper
first of all we derive the equations of circular loops of cosmic
string in the de Sitter spacetime by means of the Nambu-Goto
action with an additional factor for the time-dependent tension.
We solve the equations numerically to study the evolution of loops
and the time dependence of tension on the fate of cosmic string
loops, in particular whether the necessary conditions of larger
initial size will be revised. The discussions and conclusions are
emphasized in the end.

We start to consider the evolution of cosmic string loops whose
tensions are functions of cosmic time in a de Sitter spacetime.
The metric describing the world is written as,

\begin{equation}
ds^{2}=(1-\frac{r^{2}}{L^{2}})dt^{2}-\frac{dr^{2}}{1-\frac{r^{2}}{L^{2}}}
-r^{2}(d\theta^{2}+\sin^{2}\theta d\varphi^{2})
\end{equation}

\noindent where $L=\sqrt{\frac{3}{\Lambda}}$ is the de Sitter
radius associated with the cosmological constant $\Lambda$. A free
string propagating in a spacetime sweeps out a world sheet which
is a two-dimensional surface. The Nambu-Goto action for a cosmic
string with time-dependent tension is given by,

\begin{equation}
S=-\int d^{2}\sigma\mu(t)[(\frac{\partial
x}{\partial\sigma^{0}}\cdot\frac{\partial
x}{\partial\sigma^{1}})^{2}-(\frac{\partial
x}{\partial\sigma^{0}})^{2}(\frac{\partial
x}{\partial\sigma^{1}})^{2}]^{\frac{1}{2}}
\end{equation}

\noindent where $\mu(t)$ is the string tension and the function of
cosmic time. $\sigma^{a}=(t, \varphi)$ $(a=0, 1)$ are timelike and
spacelike string coordinates respectively. $x^{\mu}(t, \varphi)$
$(\mu, \nu=0, 1, 2, 3)$ are the coordinates of the string world
sheet in the spacetime.

For simplicity and without generality we assume that the string
lies in the hypersurface $\theta=\frac{\pi}{2}$, then the
spacetime coordinates of the world sheet parametrized by
$\sigma^{0}=t$, $\sigma^{1}=\varphi$ can be selected as $x=(t,
r(t, \varphi), \frac{\pi}{2}, \varphi)$.

In the case of planar circular loops, we have $r=r(t)$. According
to the metric (1) and the dpacetime coordinates mentioned above,
the Nambu-Goto action with an additional factor for the
time-dependent tension belonging to a cosmic string denoted as (2)
is reduced to,

\begin{equation}
S=-\int dtd\varphi\mu(t)r(h-\frac{\dot{r}^{2}}{h})^{\frac{1}{2}}
\end{equation}

\noindent giving rise to the following equation of motion for
loops which have the time-varying tension for simplicity as
$\mu(t)=\mu_{0}t^{q}$,

\begin{equation}
h^{2}r\ddot{r}+\frac{q}{t}r\dot{r}(h^{2}-\dot{r}^{2})+4hr^{2}\dot{r}^{2}
-h\dot{r}^{2}+h^{3}(1-2r^{2})=0
\end{equation}

\noindent here $h=1-\frac{r^{2}}{L^{2}}$ and $\mu_{0}$ is a
constant and we let $L=1$ here. In the case of constant tension as
$q=0$, equation (4) is reduced to what has been derived and solved
by Larsen [21]. It was shown that loops of cosmic string will keep
on expanding in de Sitter spacetimes if their radii satisfy the
condition like $r(t_{0})>0.707L$ at the moment of their formation,
or they will contract to become black holes, which means that the
initial radii of expanding loops must be large enough. Now we
solve equation (4) numerically. Having performed the burden and
surprisingly difficult calculation, we find that our conclusion is
the same as Larsen's [21] when the initial radius
$r(t_{0})>0.707L$. We discover that in the case of
$r(t_{0})<0.707L$ there must exist a critical value denoted as
$\alpha$. When $q<\alpha$, the cosmic string loops will enlarge to
evolve or contrarily will collapse to form black holes when
$q>\alpha$. Our findings indicate that the critical value $\alpha$
is negative and associated with the initial size of cosmic string
loops, which is depicted in Figure 1. The larger the initial
radius $r(t_{0})$ is, the smaller the critical value $\alpha$ is.
We also show that the cosmic string loops in the case of
time-dependent tension with $q>\alpha$ contract faster than ones
in the case of constant tension with initial radius of loops like
$r(t_{0})<0.707L$ and $q$ is not allowed to be equal to a negative
integer unless $\dot{r}(t_{0})\neq0$.

The observational results show that there could exist a lot of
cosmic string loops in our universe [18, 25]. In backgrounds with
positive cosmological constant only loops whose radii are larger
than a critical value $0.707L$ at the moment of their formations
can avoid becoming black holes no matter the strings have constant
or changeable tension. In fact the special value $0.707L$ is
extremely large and can even compares favourably with the size of
universe. As we mentioned above the larger loops must evolve to
become smaller ones copiously and continuously. The discussions in
the case of larger loops of cosmic string exhibit that only fewer
loops can survive in our accelerating universe. The conclusion
that the only sufficiently large loops of cosmic string remain
expanding in the de Sitter spacetime will not let enough cosmic
string loops exist. In the case of smaller ones we let the tension
is proportional to the power of time denoted as
$\mu(t)=\mu_{0}t^{q}$ for simplicity and without losing
generality. We find that in de Sitter spacetimes the loops with
$q<\alpha$ will expand to evolve instead of becoming black holes.
The dependence of negative critical value $\alpha$ on the initial
radius of cosmic string loops is shaped in Figure 1. When the
power $q$ is negative, the tension will be smaller and smaller and
its influence leading the loops contract become weaker and weaker.
It is clear that the final results of cosmic string loops will not
be black holes if the tensions obey the conditions that the power
is lower than a critical value. Certainly the expression of
tension can be different, but there must exist the relevant
conditions that the expression must satisfy and this kind of
cosmic string loops can remain expanding. Our findings indicate
that there may be a considerable number of loops of cosmic string
in the universe however their tension is time function obeying the
necessary conditions, which keeps our observing the effect of
cosmic string loops.

The main results of this paper is equation (4), the circular loop
equation for a cosmic string evolving in the hypersurface with
$\theta=\frac{\pi}{2}$ in the de Sitter spacetime. A larger cosmic
string loop with initial radius $r(t_{0})>0.707L$ will evolve
instead of collapsing to form a black hole no matter whether its
tension is constant. The solution to this equation shows that a
smaller loop with $r(t_{0})<0.707L$ may also never contract to one
with a Schwarzschild radius if the expression for tension of
cosmic string is a time function satisfying the necessary
condition like $\mu(t)\propto t^{q}$ and $q<\alpha$. The critical
value $\alpha$ is negative and depends on the initial size of
cosmic string loops. The larger loops need larger absolute value
of $\alpha$. Therefore a lot of cosmic string loops can evolve to
survive in our accelerating universe.

\vspace{1cm}
\noindent \textbf{Acknowledge}

This work is supported by NSFC No. 10333020 and the Shanghai
Municipal Science and Technology Commission No. 04dz05905.

\newpage
\begin{figure}
\setlength{\belowcaptionskip}{10pt} \centering
  \includegraphics[width=15cm]{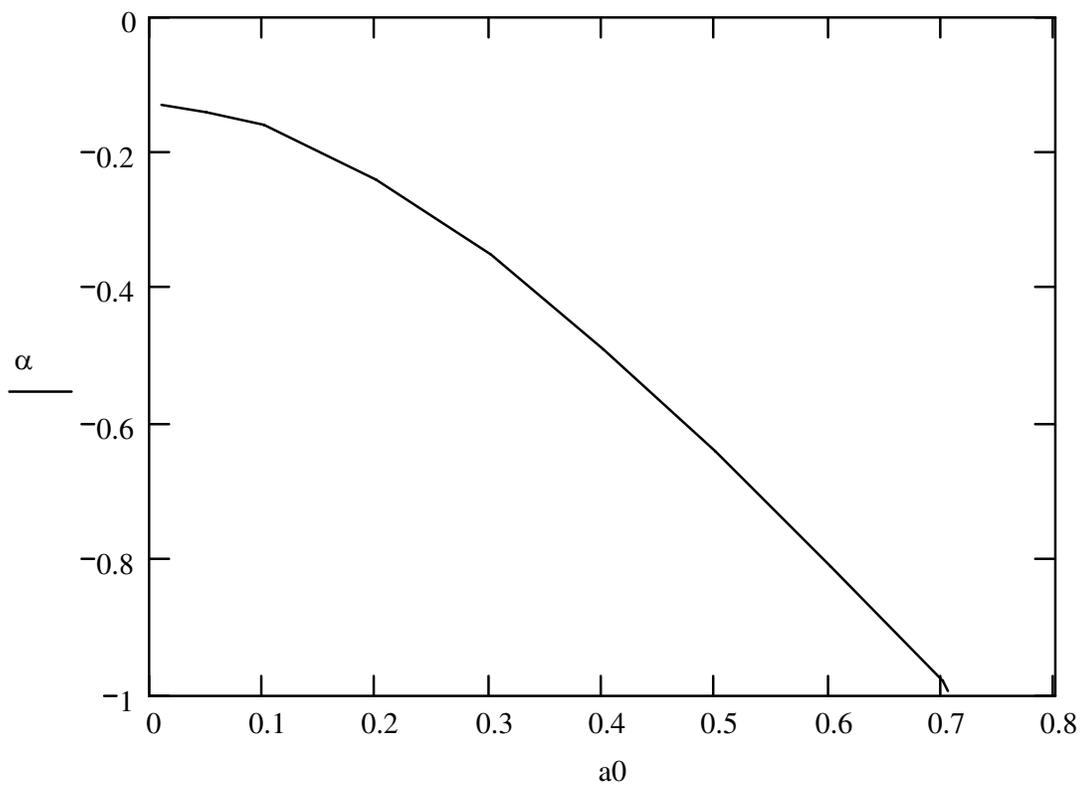}
  \caption{The curve of critical value $\alpha$ as functions of $r(t_{0})$
  initial radius of circular loops of cosmic string with $r(t_{0})<0.707L$ in de Sitter spacetime.}
\end{figure}


\newpage
\begin{thebibliography}{99}
\bibitem {Kibble} T. W. B. Kibble, Phys. Rep. 67(1980)183
\bibitem {Vilenkin} A. Vilenkin, Phys. Rep. 121(1985)263
\bibitem {Vilenkin} A. Vielenkin, E. P. S. Shellard, Cosmic
Strings and Other Topological Defects, Cambridge University Press
1994.
\bibitem {Hindmarsh} M. B. Hindmarsh, T. W. B. Kibble, Rep. Prog.
Phys. 58(1995)477
\bibitem {Pogosian} L. Pogosian, M. C. Wyman, I. Wasserman,
JCAP0409(2004)008
\bibitem {Jeong} E. Jeong, G. F. Smoot, Astrophys. J. 624(2005)21
\bibitem {Sazhin} M. V. Sazhin et. al., astro-ph/0406516
\bibitem {Sarangi} S. Sarangi, S. H. H. Tye, Phys. Lett.
B536(2002)185
\bibitem {Pogosian} L. Pogosian, S. H. H. Tye, I. Wasserman, M.
Wyman, Phys. Rev. D68(2003)023506
\bibitem {Sazhin} M. Sazhin et al., Mon. Not. R. Astron. Soc.
343(2003)353
\bibitem {Schild} R. Schild et al., Astron. Astrophys.
422(2004)477
\bibitem {Damour} T. Damour, A. Vilenkin, Phys. Rev.
D71(2005)063510
\bibitem {Hogan} C. J. Hogan, Phys. Rev. D74(2006)043526
\bibitem {Pogosian} L. Pogosian, A. Vilenkin, Phys. Rev.
D70(2004)063523
\bibitem {Olum} K. D. Olum, A. Vilenkin, Phys. Rev.
D74(2006)063516
\bibitem {Sazhin} M. Sazhin et al., MNRAS 343(2003)353
\bibitem {Sazhin} M. Sazhin et al., Astrophys. J. 636(2005)L5
\bibitem {Laix} A. de Laxi, T. Vachaspati, Phys. Rev.
D54(1996)4780
\bibitem {Anderson} M. R. Anderson, The mathematical theory of
cosmic string - cosmic strings in the wire approximation, IOP
Publishing Ltd., 2003
\bibitem {Li} X. Li, H. Cheng, Class. Quantum Grav. 13(1996)225
\bibitem {Larsen} A. L. Larsen, Phys. Rev. D50(1994)2623
\bibitem {Gu} Z. Gu, H. Cheng, Gen. Rel. Grav. 39(2007)1
\bibitem {Yamaguchi} M. Yamaguchi, Phys. Rev. D72(2005)043533
\bibitem {Ichikawa} K. Ichikawa, T. Takahashi, M. Yamaguchi, Phys.
Rev. D74(2006)063526
\bibitem {Schild} R. Schild, Astron. J. 129(2005)1225


\end{thebibliography}
\end{document}